\documentclass[aps,prl,twocolumn,groupedaddress]{revtex4}
\begin{document}

%\preprint{}

\title{Bell's Theory with no Locality assumption: putting Free Will at work.} 


\author{Charles Tresser}
\email[]{charlestresser@yahoo.com}
%\thanks{}
\affiliation{IBM, P.O.  Box 218, Yorktown Heights, NY 10598, U.S.A.}


\date{\today}

\begin{abstract}
We prove a version of the Bell's Theorem that does not assume Locality but only the Effect After Cause Principle (EACP) according to which for any Lorentz observer the value of an observable cannot change because of an event that happens after the observable is measured. Since the EACP is compatible both with Locality and with Non-Locality, Locality cannot be considered as the common cause of the contradictions obtained in all versions of Bell's Theory.  By definition, all versions of Bell's Theorem assume Weak Realism according to which the value of an observable needed in the discussion of Bell's Theorem is well defined whenever the measurement could be made and some measurement is made. As a consequence of our results, Weak Realism becomes the only hypothesis common to the contradictions obtained in all versions of Bell's Theory.  This work indicates that it is Weak Realism, not Locality, that needs to be negated to avoid the contradictions in microscopic Physics associated to Bell's Theory, at least if one refuses as false the de Broglie-Bohm Hidden Variable theory because of its essential violation of Lorentz invariance.  This paper completes with much more details the genuine Bell Theorem part of a previous paper.  That paper also offered a treatment of the GHZ entanglement, a treatment which did not suffer from the lack of clarity of the definition of the EACP.
\end{abstract}

\pacs{03.65.Ta}
%\keywords{}

\maketitle

%\section{}
%\subsection{}
%\subsubsection{}
%
%
%
%

\noindent
\textbf{1)}  \textbf{Introduction.}

\emph{\bf This paper is dedicated to Pierre Coullet, my big brother with whom I did my most important contributions to science, or at least to macroscopic physics after we dreamed together for years about collaborating and about having the chance to make a career in research.  I may say that without the many way Pierre helped me, I would probably never have had such an opportunity despite my childhood dreams.  This is why I want to dedicate to Pierre the cleanest version of what is so far my main contribution to microscopic physics, one of the rare papers of which I have ever been proud.  After all microphysics is the branch of sciences about which Pierre and I most often dreamt together as young adults. There are probably not many odd places were we have not spoken ardently about physics together and it has been a great pleasure recently to collaborate again on a few projects. Has such craziness come to an end for him and I because of old age: I would not bet on it.  Pierre hates birthdays, and why would he like getting old officially while he is still as ardent and filled with energy and new great ideas (most people can be filled with small or bad ideas) as he ever was?  Recently, Pierre has turned part of his attention to classical optics (and other things), about at the time when I got sucked into the crazy world of the foundations and interpretation of Quantum Mechanics, where the experiments and many of the contributors are in Quantum Optics. In fact there has often been parallels between our evolutions in the years when we did not write any joint paper (not so parallel as converging since we rewrote papers together after all). I hope that Pierre will present us with many more years of brilliant ideas and great achievements, whatever the domains of human activity he wishes to contribute to, wherever he wants to spend his formidable energy.}

The present paper completes significantly the first part, concerning the genuine Bell Theorem, of a paper entitled \emph{Bell's Theory with no Locality assumption} \cite{TresserAutre}, whose second part deals with the so called GHZ entanglement. To make the present treatment of Bell's Theory fully independent from the former version, everything that could be needed for the present discussion is repeated here so that the readers will never need to have both papers at once in their hands.  To the contrary whatever concerns GHZ \cite{GHZ1989, Mermin1990GHZ3, GHSZ1990GHZ3} in the first paper is not even sketched here. In between the two papers, I have come to realize that it was needed to make fully explicit that what I call the EACP is fully compatible with whichever of Locality or Non-locality that one may choose as a supplementary hypothesis to EACP. This makes the proof of one essential lemma disputable, which I correct, but also this better understanding of the EACP permits a much clearer presentation, of which reading the GHZ part of \cite{TresserAutre} after the present paper would greatly profit. A good usage of these two papers would thus consist in first reading this paper in full and then reading the GHZ part of \cite{TresserAutre}.

Following Bohm 's version \cite{Bohm} of the EPR \textit{gedanken experiment} \cite{EPR}, we consider entangled pairs of spin-$\frac{1}{2}$ particles such that the spin part of the wave function is the \emph{singlet state} (at any  pair of locations $(x_1,x_2)$):
\begin{equation}\label{Singlet}
\Psi(x_1,x_2)=\frac{1}{\sqrt{2}}(| +\rangle _A\otimes| -\rangle_B-| -\rangle_A\otimes|
+\rangle_B)\,.
\end{equation}
The sum of tensor products that we see in (\ref{Singlet}) represents an example of \emph{entanglement}, which means that this expression cannot be rewritten as one tensor product of one particle states.  We even have here a \emph{maximal entanglement} since all the summands have identical statistical weights.  The particles of the pair indexed by $i$ are called $(p_A)_i$ and $(p_B)_i$.  For each $i$ the particle $(p_A)_i$ flies to Alice who is equipped with the \emph{measurement tool} $E$ while $(p_B)_i$ flies to Bob who handles the \emph{measurement tool} $P$.  $E$ and $P$ can be chosen as Stern-Gerlach magnets if, following Stapp \cite{Stapp1971} we use  neutral particles, say neutrons.  There is a source  $S$ of entangled pairs and, together with the tools $E$ and $P$, the source $S$ is attached to the laboratory frame; we assume that the measurements at $E$ and $P$ are (essentially) simultaneous in that frame.  

With the successive values of $i$ associated to successive pairs of entangled particles both of which get detected when needed, Alice chooses the oriented axes $(a_A)_i$ and, with $ \textrm{Spin}(q)$ standing for the spin of particle $q$, she  observes the sequence $\mathcal E_i$ of normalized projections of $ \textrm{Spin}((p_A)_i)$ along $(a_A)_i$ while Bob chooses the oriented axes $(a_B)_i$ and observes the sequence $\mathcal P_i$ of normalized projections of $ \textrm{Spin}((p_B)_i)$ along $(a_B)_i$.  

Bohm \cite{Bohm} noticed in particular that any observation $\sigma \in \{-1,+1\}$ by Alice along $(a_A)_i$ would necessarily correspond, because of the structure of the singlet state, to the observation $-\sigma$ by Bob if he would choose $(a_B)_i=(a_A)_i$.  We will not recall nor revisit here the EPR paper, nor comment the way the authors themselves or Bell considered the issues raised in \cite{EPR} or in \cite{Bohm}.  The consideration of angles between the oriented axes  $(a_A)_i$ and  $(a_B)_i$ that can take any value in a setting that is otherwise the one proposed by Bohm is essential in the development of Bell's Theory \cite{Bell}, and it is precisely this theory that we revisit here (Bohm only used right or zero angles between axes in \cite{Bohm} as he was merely proposing a new version of what he considered to be the content of the EPR paper \cite{EPR}).  The experiment that consists in emitting successive pairs in the singlet state and measuring the normalized projections of the associated spins is repeated a large number of times in order to get statistically significant results. To help in achieving the same goal of significant statistics, the oriented axes $(a_A)_i$ and $(a_B)_i$ are usually kept constant for long sequences of values of $i$ (on such stretches of constancy, one may prefer to suppress the indexation of the axes).  If the observation is $\mathcal Q_i$, we write $|\mathcal Q_i\rangle$ for the corresponding spin part of the state, and we denote by $X$ the sequence with generic element $X_i$ (\textit{i.e.,} we use an index together with the symbol for a sequence to designate a special or a generic element of that sequence).  Still about concepts and notations that we will use, we mention:
 
 - The \emph{correlation} 
 \begin{equation}
 \langle \mathcal U,\mathcal V\rangle =\lim_{N\to\infty}\frac{1}{N}(\mathcal U_{i+1}\cdot \mathcal V_{i+1}+\dots+\mathcal U_{i+N}\cdot \mathcal V_{i+N})\,,
 \end{equation}
this limit being equal to Dirac's braket $\langle \mathcal U|\mathcal V\rangle$ whenever both  $|\mathcal U\rangle$ and  $|\mathcal V\rangle$ are quantum mechanical states, but we will prefer the statistical notation that allows one to use liminf and lim sup when we are no more inside the Realm of Quantum Mechanics and the convergence to a limit is no more guarantied so that the correlation gets then a lower and an upper bound rather than a definite value. 
 
 - The \emph{angle} $<a_1;a_2>$  between the two oriented axes $a_1$ and $a_2 $ that lets us associate to any oriented axis $a_C$ the angle $\theta_C=<a_0;a_C>$  where $a_0$ is some oriented axis of reference that points, say, horizontally and to the right when looking from the far side along the estimated classical trajectory of the departing particle flying toward Alice.  

\noindent 
Recall that for the quantities of interest here, Quantum Mechanics predicts probabilities of equality or equivalently correlations, the equivalence of the two viewpoints being captured in our case in the following identity:
\begin{equation}\label{CorrelProba}
\langle  \mathcal E,  \mathcal P \rangle= 2 Prob( \mathcal E_i= \mathcal P_i)-1
\end{equation}
where $Prob($event$)$ is the probability of that $event$ (for a general reference for Quantum Mechanics covering somehow Bell's Theory and further entanglements, see for instance \cite{Peres1993} or \cite{Le Bellac}).

\medskip
For the sequences $ \mathcal E$ and $ \mathcal P$ that we have defined for the  spin-$\frac{1}{2}$ singlet state (\ref{Singlet}), Quantum Mechanics predicts what we call the \emph{twisted Malus Law} that differs from the usual Malus law by the minus sign:
\begin{equation}\label{TwistedMalus}
\langle   \mathcal E,    \mathcal P \rangle= -\cos (\theta_A-\theta_B)\,.
\end{equation}
Since we only use spin-$\frac{1}{2}$ particles and normalized spin projections rather than photons and their polarization states, each time we mention in this paper the singlet state or Malus law (normal or twisted), we mean of course the spin-$\frac{1}{2}$ version of these objects (for a textbook presentation of both of the photons and the  spin-$\frac{1}{2}$ particles versions, see for instance \cite{Le Bellac}).  

\medskip
We now turn to the subject matter of the present paper and first recall that in the founding paper of Bell's Theory \cite{Bell} (see page 407 of that paper), Bell reached the conclusion that:

\smallskip
\emph{``In  a theory in which parameters are added to quantum mechanics to determine the results of individual measurements, without changing the statistical predictions, there must be a mechanism whereby the setting of one measuring device can influence the reading of another instrument, however remote.  Moreover, the signal involved must propagate instantaneously, so that such a theory could not be Lorentz invariant."}.

\bigskip
\noindent
More generally, the structure of a typical Bell type theorem reads either as the following statement that we call \emph{the Main Implication} or as its consequences as in Bell's citation just above:

\medskip
\noindent
\textit{ \,Quantum Mechanics} \quad \quad \quad \textit{Some inequality is violated }

\noindent
\textit{+ Augmentation choice} \,\,\,$ \Rightarrow \,\,$\textit{ for appropriate choices} 

\noindent
\textit{+ Extra hypothesis}\qquad \qquad \, \textit{ of some parameters.}

\medskip
\noindent
In the terms of the Main Implication, the example of ``Augmentation" chosen in Bell's 1964 paper \cite{Bell} is the assumption that there are ``Predictive Hidden Variables with the same statistics as Quantum Mechanics" while Bell's original example of ``Extra hypothesis" is ``Locality" that we next redefine both more formally and in such a way that the role of the augmentation be clearly stated.    

\medskip
\noindent
\textbf {Definition 1.}  \emph{Locality tells us that if $(x_0, t_0)$ and $(x_1, t_1)$ are spatially separated, \textit{i.e.,}  $\Delta x ^2> c^2 \Delta t^2$, with $c$ standing for the speed of light,  then  the output of a measurement made at  $(x_1, t_1)$ cannot depend upon the setting of an instrument at  $(x_0, t_0)$ (making no measurement at $(x_0, t_0)$ being in this context one possible apparatus setting). Furthermore, if one assumes that Weak Realism as defined below holds true,  the value of any observable that could be measured at $(x_1, t_1)$ in lieu of the observable that is actually being measured there is also independent of any instrument setting at $(x_0, t_0)$.}

\smallskip
\noindent
In the present paper we will show that the Extra hypothesis of Bell's Theorem can be chosen to be substantially weaker than Locality without affecting the truth of the Main Implication: the price to pay for weaker hypothesis will be a much reduced set of situations supporting a Bell Theorem without a Locality assumption.  

\medskip
For a long time already many authors have proposed versions of Bell's Theorem based on Augmentations that are weaker than the Predictive Hidden Variables used in \cite{Bell}.  We recall that the concept of \emph{Predictive Hidden Variables} does not only mean that some variables make sense, even if beyond our reach, but that there are enough such variables so that using all variables, hidden or not, one would get a theory that would not only predict statistical results (like Quantum Mechanics) but would also predict the result of individual experiments and more generally of all the observables' values (even if  one cannot access these values). In particular all usual observables would have well defined values since they would be predictable, so that Predictive Hidden Variables, if they would exist, would imply \emph{Realism} (also called \emph{Microscopic Realism}, for instance in \cite{Leggett2008} ) in the sense that observables would have values (possibly unknown and that possibly cannot be known) independently of being observed or not.  We shall focus in this paper on an Augmentation of Quantum Mechanics that does not \emph{assume} any more predictive power than Quantum Mechanics; more precisely we shall only postulates the following augmentation:

\medskip
\noindent
\textbf{Weak Realism.} \emph{There is a value associated to any {\bf useful} measurement that could be made on a particle at the time when some measurement is made on that particle, where the usefulness qualification is attributed to an observable if it is used to formulate a Bell Theorem. Furthermore, it is assumed that the values of the observable that exist according to Weak Realism preserve the statistical predictions of Quantum Mechanics.}

This Augmentation is mostly Stapp's  \emph{Contrafactual Definiteness} \cite{Stapp1985} (see also \cite{Stapp1971}), also called \emph{Macroscopic Counterfactual Definiteness} and abbreviated as MCFD by Leggett who also offers an interesting discussion of it in \cite{Leggett2008}, is not only implied by the hypothesis used by Bell in \cite{Bell} but also by Microscopic Realism as explained in \cite{Leggett2008} and also by what is called sometimes \emph{the EPR  condition of reality} \cite{EPR}.  

\medskip
Thus the Augmentation of Quantum Mechanics that we choose in this paper in order to develop a restricted hypotheses version of Bell's Theory is the small modification of MCFD that is stated above. We have added the word {\bf ``useful"} in the usual definition of MCFD so that we can recognize that, by definition, Weak Realism is constructed as:

- \emph{The weakest form of realism sufficient to develop Bell's Theory}  

\noindent
while 

- \emph{Preserving the statistical predictions of Quantum Mechanics (by a standard important extra hypothesis going back to\cite{Bell}: see also Convention 2 below).}

\noindent
This is why we call the chosen Augmentation simply \emph{Weak Realism} (``Weakest Realism" would have been more precise but seems less appealing). At first reading, the readers may as well choose the formulations of realism at the microscopic level that they like most as the definition of Realism, in lieu of our minimalist concept. We will say \emph{``Realism"} to mean \emph{``any form of Realism"}, and whenever \emph{``Weak Realism"} is invoked,  \emph{``any (reasonable) form of Realism"} could be used instead.

\medskip
The concept that we introduce next is another essential ingredient of our work: it will be our Extra Hypothesis in the Main Implication.    

\medskip
\noindent
\textbf {Effect After Cause Principle (\emph{EACP} - General Form):} \emph{- (i) For any Lorentz observer the value of an observable cannot change as a result of any cause that happens after said observable has been measured for that observer.}

\noindent
\emph{- (ii) Furthermore, if one assumes that Weak Realism holds true,  the value of any observable that could be measured at $(x, t)$ where some other observable is measured, but that is only inferred to exist at $(x,t)$ by invoking Weak Realism, cannot change as a result of any cause that happens after said non-observed observable gets a value at $(x,t)$ as a result of Weak Realism for that observer.}

\bigskip
We will mostly use the EACP in a form that is much more specific that the one proposed here (see Subsection 3.1) but we have judged that it was preferable to stay at this less technical level in the Introduction, and we will also use this high level description anyhow. The subtlety of (both of) the statement(s) of the EACP calls for some special comments that are more of the ``warning" type than ordinary remarks.

\medskip 
\noindent
\textbf{Warnings.} \emph{
\begin{description}
\item[W1] We notice that time ordering used in the definition of the EACP is relative to the chosen Lorentz observer. 
\item[W2] Accepting that Locality fails to always hold true means that the value of an observable may well depend on a ``later" event in the time ordering of the chosen Lorentz observer: see Warning [W1] : the next warning [W3] provides limits for the dependence on later events. 
\item[W3] A reading of an observable, once performed (or potentially performed when dealing with entities that only exist if one assumes Weak Realism) cannot (further) change because of a later cause, but the reading may \emph{depend} or \emph{not depend} on a cause that happens later for some Lorentz observer according to whether one assumes Non-Locality or Locality.
\end{description}
} 

\medskip
\noindent
\textbf {Free Will Principle.} We will also use the \emph{Free Will Principle}, (or  \emph{FWP}) which (in the rather minimal version that we choose) states that given an experimentalist at time $t_0$ and a neighborhood \textbf{N}$_t$ of this experimentalist for times $t$ close to but smaller than $t_0$ (the neighborhoods in three-space  \textbf{N}$_t$ being chosen so as to overcome the problem that while the experimentalist is not punctual, we still want a neighborhood that contains a full experimentalist at some time equal or close to $t_0$  and earlier version of the same person for some Lorentz observer so that all the parts of the experimentalist can participate to a decision), then no event out of the union of the closed past light cones of the points in $\textbf{N} =\bigcup _t\,\textbf{N}_t$ should affect (neither directly nor with the mediation either of parts of  \textbf{N} or of the past light cones of such parts of \textbf{N}) the ability of the experimentalist to decide what she or he wants to do in particular in terms of not using or using and then in terms of how to use some apparatus relevant in an experiment that has been set up in the Bell's Theory context.  

\noindent
The version of free will that we use is quite weak because we did not want to get into the issue of whether freedom of will contradicts \emph{determinism}, an issue particularly important here as realists tend to defend determinism together with realism and sometime defend realism mostly in order to save determinism in the sense of Laplace. With such a weak form of the FWP, if Alice is a machine, we could still ask that said machine has some form of free will but we will refrain from getting into the machine's freedom issue or any related deep issue. 

\noindent
It may seem that the FWP, even in the weak sense that we use, contradicts Non-Locality in the way it is used in the context of Bell's theory.  However, the way Non-locality comes in usually is by allowing new sequences, either measured or resulting from Weak Realism, when the axes used by Alice or Bob are changed: the low impact of such changes -that explains why SMT does not result from Non-locality- relates only to correlation between pairs of axes with one axis for $E$ and one for $P$: assuming Non-Locality the pairs $(E,E')$ and $(P,P')$ get harder to treat as we have seen above, but that is definitely unusual utilization of the impact of Non-Locality.  Indeed we will see that in one very special case, the FWP could create conflicts with the EACP if one imposes that Non-Locality affects correlation values for expressions such as $\langle{\mathcal E}, {\mathcal E}' \rangle$ and not only the respective schedules of outputs at $E$ and $P$. This state of affair does not change the fact that Free Will does nor preclude Non-Locality, a typically low impact phenomenon, and Non-Locality would probably not have survived such an incompatibility for many physicists anyway.

\noindent
We could possibly have used a version of the \emph{Free Will Theorem} (or \emph{FWT}) \cite{FreeWill1} that would deal with two spin-$\frac{1}{2}$ particles rather than with two spin-1 particles as in \cite{FreeWill1} instead of using directly the FWP.  However we did not seriously try the FWT path since one the one hand the argument on the basis of the FWP became clear soon enough while on the other hand we could not locate extra value that following the FWT path would bring, assuming it viable, not even mentioning that we are not yet sure that we could find a two spin-$\frac{1}{2}$ particles version. This gap is possibly a good direction for future investigations, but it might be preferable to rather look for an EACP based treatment of GHZ \cite{GHZ1989, Mermin1990GHZ3, GHSZ1990GHZ3} as proposed in \cite{TresserAutre} or of the two spin-1 particles configuration considered in \cite{FreeWill1}: the treatment of the GHZ case reported in  \cite{TresserAutre} was first feared to be incomplete and indeed feared to call at least for a much more detailed discussion
while in fact it turns out that the only impact of the clearer version of the EACP that we use only invites us to call ones attention to the fact that the algebra remains unchanged when one replaces locality by the EACP in the GHZ context.  Anyway, the version of Free Will that we use here (in order to avoid conflicts that are inessential to our main goal) is much weaker than the version in  \cite{FreeWill1} that assumes a failure of determinism.

\bigskip
After showing that the EACP is an hypothesis different from and not stronger than Locality, and offering a discussion a bit less formal indicating that the EACP is indeed weaker an hypothesis than Locality, we will prove a version of Bell's Theorem where we only assume Weak Realism and  the EACP.  Let us recall that a typical formal statement of Bell's Theorem consists in the \textit{falsification of some inequality} (meaning as of old the exhibition of an instance such that the inequality that one attempts to \textit{falsify} indeed reduces to a false inequality between two numbers): such a falsification then lets one draw a conclusion as in Bell's citation reported above.  Not so surprisingly, there is a price to pay for weakening the hypothesis of the new Bell's Theorem that we will prove in this paper; more specifically, in order to compensate for our weaker assumptions, the selection of an inequality and of its parameterization needs to be much more controlled than in former versions of Bell's Theory in order to produce a falsification of at least one of the Bell's Inequalities than what one needs when as usual on assumes Locality.  In particular, avoiding to assume Locality will not allow us to falsify  any inequality that uses projections of the spin along two axes for each of the two particles of the singlet state (so four axes, or if one prefers, four angles) as in the CHSH version of Bell's Theorem (see \cite{CHSH69, Bell71, CH74, AspectEtAl1982}). However the original configuration with three angles (two on one side, one one the other particle for each pair) used in Bell's paper \cite{Bell} can be dealt with, but then only when the axes are at a right angle of each other on the side where one uses two oriented axes when one only assumes the EACP and Weak Realism.  It might be the case that other configurations also work besides the one that we could find, but the fact that we cannot conclude using two angles for each of the two particles of the singlet state will turn out to be deeply linked to the difference between Locality and the EACP.

\medskip
We conclude from our results that \emph{the only  cause that is common to all contradictions in Bell type theorems is whatever form of realism that one uses to augment Quantum Mechanics}.  In particular, invoking Non-Locality cannot prevent the contradictions that we establish since we do not assume Locality to begin with.  Theorems of the type of Bell's can thus be used to strongly suggest that \emph{ Weak Realism is false}  (see also  \cite{ETP, Tresser2} and Remark 3 below).  Such a conclusion would be in line with the opinion that any form of realism at the microscopic level is in contradiction with the spirit of the Uncertainty Principle \cite{Heisenberg} .  Rejecting as usual ``Local Realism", \textit{i.e.,} the conjunction of some form of realism and Locality, appears to be a misleading conclusion in view of our results. 

\bigskip
This paper is organized as follows. In Section 2 we complete the description, started above in this section, of the setting of the \emph{gedanken experiments} dealt with in Bell's Theory: for the sake of completeness, simple classical derivations of Bell's Inequality and Bell's Theorem, in the original  and CHSH form, are provided there assuming both Weak Realism and Locality as in the classical Bell's Theory.  In Section 3, we prove that the EACP is an hypothesis that is neither Locality nor weaker than Locality and we further compare these two hypothesis.  We will also provide in Section 3 a computation of one special example of correlation that we call the \emph{No Correlation Lemma}.  The proof of our version of Bell's Theorem is then developed to conclude Section 3. 

\bigskip
\noindent
\textbf{2)}  \textbf{Setting, statement and Proofs in usual Bell's Theory.}
We started the description of the experiment in the Introduction, but what was described there of the EPR-Bohm \textit{gedanken experiment} is not yet enough to reach any form of Bell's Theorem. The extended version of the Uncertainty Principle \cite{Robertson} tells us that only one spin projection, \textit{i.e.,} one axis, can be chosen for each of the two particles $p_A$ and $p_B$, not enough to generate any meaningful inequality relating different correlations. In order to get enough observables to build a meaningful inequality, one needs to ``augment" Quantum Mechanics into a candidate for a theory of microphysics that would coincide with Quantum Mechanics where Quantum Mechanics has something to tell us, and that is compatible with the statistical predictions of Quantum Mechanics (which have been proven right by numerous experiments over the years).  Using the Augmentation of Quantum Mechanics by any form of realism to have more values of observables at once necessarily turns what we started to describe as an experiment into a \textit{gedanken} experiment.  Of course, the legitimacy of such an augmentation of Quantum Mechanics is questionable and we hope that we help to make the case (see Remark 3) that indeed,  Weak Realism violates the laws of Physics.  But this will not prevent us from often assuming Weak Realism as we argue \textit{ad absurdum}.  

\medskip
The following two conventions are adopted in a more or less explicit form in all works on Bell's Theory, independently of the strength of the Augmentation being chosen:

\bigskip
\noindent
\textbf{Convention 1.}  \emph{Whenever we assume that Quantum Mechanics is augmented by a form of realism, we implicitly postulate that \emph{any quantity that is not measured but that exists according to the Augmentation has the value that would have been measured if this quantity would have been the one measured, the world being otherwise unchanged}.  It seems to us that the meaning of the value of an observable makes no much sense otherwise so that this convention is probably the most consensual component of this paper. }

\medskip
\noindent
\textbf{Convention 2.}  \emph{Whenever we assume that Quantum Mechanics is augmented by a form of realism, we assume that said Augmentation is made \emph{without changing the statistical predictions}.  This is (up to wording) the assumption that Bell made in his foundational 1964 paper \cite{Bell}, except for the fact that we do not restrict the choice of Augmentation to Predictive Hidden Variables.}

\medskip
As we shall see, Convention 2 (that extend one assumption of Weak Realism to any form of Realism useful in our context) is not enough to get convergence, nor even evaluation for averages over finite sums for all the correlations that we need.  Historically, versions of inequalities involving either three sequences of spin projections (what we call \textit{``version $V3$"}) or four sequences of spin projections (what we call \textit{``version $V4$"}) have been used, and it will be important for us to use both versions.   More precisely, we will use: 

- Version $V3$ in order to get our Bell Theorem without Locality  in subsection 3.3.  

- Versions $V3$ and $V4$ in order to illustrate why the EACP is a weaker hypothesis  than Locality in subsection 3.1.
\noindent
We will also use both versions $V3$ and $V4$ to examine closely in Remark 2 of subsection 3.1 what would be the cost of abandoning Locality that is used as an essential assumption when dealing with the usual Bell's Theory and replacing it by the EACP.  

\medskip
Coming back to the setting and notations of the Introduction, and assuming Weak Realism so that extra axes $(a_A)'_i$ and $(a_B)'_i$ can respectively be chosen by Alice and Bob, we end up having at our disposal the following sequences:  

- The sequences $\mathcal E_i$ on Alice's side using axes $(a_A)_i$ and  $\mathcal P_i$ on Bob's side using axes $(a_B)_i$ are the two sequences of normalized spin projections that are actually observed. 

- The sequences $\mathcal E'_i$ on Alice's side and (if needed)  $\mathcal P'_i$ on Bob's side that are the two sequences of what would be supplementary normalized spin projections, with values that are most probably out of reach. Such supplementary normalized spin projections values would be well defined - even if out of possible knowledge - if and only if  Weak Realism or some stronger form of Microscopic Realism holds true.  These sequences are supposedly what one would get respectively along the axes  $(a_A)'_i$ and $(a_B)'_i$ if those axes would be used to measure normalized spin projections instead of the axes $(a_A)_i$ and $(a_B)_i$.  Even if such supplementary sequences of normalized spin projections cannot be known, one may construct out of them some objects with statistical significance such as correlations or probabilities of equality on which one has grip under the standing assumption that whichever form of Microscopic Realism that one invokes must respect the statistical predictions of Quantum Mechanics.  

\medskip
One may think that the range of the index $i$ is cut into disjoint intervals $I_\kappa$ so that for any $I_\kappa$  the axes $(a_A)_i, (a_B)_i, (a_A)'_i, (a_B)'_i$ do not vary  with $i$ as long as $i$ stays in $I_\kappa$: we shall denote by $N_\kappa$ the number of elements of  $I_\kappa$, and will avoid the $\kappa$ subscript whenever possible. All the sequences that we have introduced are sequences of normalized spin projections for spin-$\frac{1}{2}$ particles, hence sequences of $-1$'s and $1$'s.  We shall next focus on abstract sequences and finite chunks of sequences of $1$'s and $-1$'s.  

\bigskip
\noindent
\textbf{2.1: The formal aspects of Bell's Inequalities:}
We now follow Sica \cite{Sica1, Sica2} (except that we defer deciding which quantities ``need a prime", something which will make sense a bit below).  Sica noticed that if $w_i$, $x_i$, $y_i$ and $z_i$ are four sequences with values in the set $\{-1,1\}$, then one has simple factorization identities that  lead \textit{via} simple algebra to inequalities involving either three or four sequences or finite chunks of these sequences. For version $V3$, we use the fact that $y_i^2\equiv 1$ to start with:

\begin{equation}\label{Sica_3_1}
x_iy_i-x_iz_i=x_iy_i(1-y_iz_i)
\end{equation}
so that by summing over the elements of  $I_\kappa$, dividing by  $N_\kappa$ and taking absolute values, we get:
\begin{widetext}
\begin{equation}\label{Sica_3_2}
|\sum_{i\in I_\kappa}\frac{x_iy_i}{ N_\kappa}-\sum_{i\in I_\kappa}\frac{x_iz_i}{ N_\kappa}|\leq\sum_{i\in I_\kappa}\frac{|x_iy_i|\cdot|1-y_iz_i|}{ N_\kappa}\leq 1-\sum_{i\in I_\kappa}\frac{y_iz_i}{N_\kappa}\,.
\end{equation}
 \end{widetext}
 Thus 
\begin{equation}\label{Sica_3_3}
|\sum_{i\in I_\kappa}\frac{x_iy_i}{ N_\kappa}-\sum_{i\in I_\kappa}\frac{x_iz_i}{ N_\kappa}|\leq 1-\sum_{i\in I_\kappa}\frac{y_iz_i}{N_\kappa}\,.
\end{equation}
Assume then that there is convergence as $ N_\kappa\to\infty$.  Denoting by $\langle f,g\rangle$ the correlation of two functions $f$ and $g$, we get: 
\begin{equation}\label{Sica_3_4}
| \langle x,y\rangle -\langle x,z\rangle|\leq 1-\langle y,z\rangle\,,
\end{equation}
one formal form of the $V3$ version of Bell's Inequalities. 

\smallskip
We now turn to the algebra of the version $V4$. Again following Sica we start with:

\begin{equation}\label{Sica_4_1_a}
x_iy_i+x_iz_i+ w_iy_i-w_iz_i=x_i(y_i+z_i)+w_i(y_i-z_i)\,.
\end{equation}

\bigskip
\noindent
Simple manipulations on this identity then yield:
\begin{widetext}
\begin{equation}\label{Sica_4_2}
|\frac{1}{N_\kappa}\sum_{I_\kappa}x_iy_i+\frac{1}{N_\kappa}\sum_{I_\kappa}x_iz_i|
+
|\frac{1}{N_\kappa}\sum_{I_\kappa}w_iy_i-\frac{1}{N_\kappa}\sum_{I_\kappa}w_iz_i|
\leq
\frac{1}{N_\kappa}\sum_{I_\kappa}|x_i|\cdot|y_i+z_i|+\frac{1}{N_\kappa}\sum_{I_\kappa}|w_i|\cdot|y_i-z_i|
\end{equation}

Now, since $\min(|y_i+z_i|, |y_i-z_i|)=0$ and  $\max(|y_i+z_i|, |y_i-z_i|)=2$, equation (\ref{Sica_4_2}) can be rewritten as

\begin{equation}\label{Sica_4_3}
|\frac{1}{N_\kappa}\sum_{I_\kappa}x_iy_i+\frac{1}{N_\kappa}\sum_{I_\kappa}x_iz_i|
+
|\frac{1}{N_\kappa}\sum_{I_\kappa}w_iy_i-\frac{1}{N_\kappa}\sum_{I_\kappa}w_iz_i|
\leq 2
\end{equation}
\end{widetext}
Assuming convergence, the averages generate correlations and one obtains the following form of the CHSH inequality, our $V4$ version of Bell's Inequalities
\begin{equation}\label{Sica_4_4}
 |\langle x,y\rangle +\langle x,z\rangle| + |\langle w,y\rangle -\langle w,z\rangle|  \leq 2\,,
\end{equation}
which contains equation (\ref{Sica_3_4}) as a special case (first restrict to $x=y$, replace each $x$ by $y$ and then rename $w$ to $x$). 

\medskip
We notice that when two sequences are actually observed so that elements with the same index come from the same pair,  then Quantum Mechanics provides the value of the correlation and in particular guaranties convergence. Notice that we made no attempt to deduce all the Bell's inequalities, formal or not. For that and the statistical aspects of Bell's Inequalities and Bell's Inequalities as a particular case of Boole's Inequalities, see for instance \cite{Fine1982a, Fine1982b, Boole1862, Pitowsky1989a, Pitowsky1991, Pitowsky1994, Pitowsky2001}.

\bigskip
\noindent
\textbf{2.2: From formal inequalities to Bell's Inequalities and Bell's Theorem:}
We have obtained versions $V3$ and $V4$ of Bell's Inequalities using abstract sequences of $1$'s and $-1$'s.  In order to come one step closer to Physics, we first appropriately pair: 

- the symbols $\mathcal E_i$,  $\mathcal P_i$ that represent actual observations, 

- and the symbols $\mathcal E'_i$ and  $\mathcal P'_i$ that represent values provided by the Weak Realism assumption 

\medskip
\noindent
to the sequences  $w_i$, $x_i$, $y_i$, $z_i$ used in deriving the inequalities (\ref{Sica_3_3}) and  (\ref{Sica_4_3}).  

\medskip
For version $V3$, we need to take $x_i$ and $z_i$ on the same side, \textit{e.g.,} Alice's side: thus $x_i=\mathcal E_i $ and  $z_i=\mathcal E'_i $, whence $y_i=\mathcal P_i $. Then equations (\ref{Sica_3_3}) and  (\ref{Sica_3_4}) become respectively:

\begin{equation}\label{Sica_3p_3}
|\sum_{i\in I_\kappa}\frac{\mathcal E_i\mathcal P_i}{ N_\kappa}-\sum_{i\in I_\kappa}\frac{\mathcal E_i\mathcal E'_i}{ N_\kappa}|\leq 1-\sum_{i\in I_\kappa}\frac{\mathcal P_i\mathcal E'_i}{N_\kappa}\,.
\end{equation}
and
\begin{equation}\label{Sica_3p_4}
| \langle \mathcal E,\mathcal P\rangle -\langle \mathcal E,\mathcal E'\rangle|\leq 1-\langle \mathcal P,\mathcal E'\rangle\,,
\end{equation}

\medskip
As for version $V4$, we want $w$ and $z$ to be the values generated by the  Weak Realism hypothesis, but  we need also $x_i$ and $z_i$ to be on differente sides and  $y_i$ and $w_i$ to be on differente sides. One way to achieve that is to choose the replacements
$x\to  \mathcal E$, $y\to  \mathcal P$, $w\to  \mathcal E'$, $z\to  \mathcal P'$. Thus equations (\ref{Sica_4_3}) and  (\ref{Sica_4_4}) become respectively:
\begin{widetext}
\begin{equation}\label{Sica_4p_3}
|\frac{1}{N_\kappa}\sum_{I_\kappa}\mathcal E_i\mathcal P_i+\frac{1}{N_\kappa}\sum_{I_\kappa}\mathcal E_i\mathcal P'_i|
+
|\frac{1}{N_\kappa}\sum_{I_\kappa}\mathcal E'_i\mathcal P_i-\frac{1}{N_\kappa}\sum_{I_\kappa}\mathcal E'_i\mathcal P'_i|
\leq 2
\end{equation}
\end{widetext}
and the following form of the CHSH inequality:
\begin{equation}\label{Sica_4p_4}
 |\langle \mathcal E,\mathcal P\rangle +\langle \mathcal E,\mathcal P'\rangle| + |\langle \mathcal E',\mathcal P\rangle -\langle \mathcal E',\mathcal P'\rangle|  \leq 2\,.
\end{equation}
Our first goal in this section is to reach the classical Bell's Inequalities and Bell's Theorems under the usual hypothesis. We also want to inspect here the correlations that can be computed if one assumes Weak Realism and Locality. This examination of what is computable under these hypotheses will be revisited in the next section where we compare the strengths of different hypotheses.

\medskip
We have already invoked Weak Realism in order to give meaning to three spin projections at once in version $V3$, or four spin projections at once in version $V4$. In order to give meaning to the correlations in equations (\ref{Sica_3p_4}) and (\ref{Sica_4p_4}), we now further assume Locality, so that the sequences on one side do not depend on the choice of the axes along which the spin is projected on the other side (no dependence on the setting of an apparatus one the other side). Then, under Conventions 1 and 2 that are both triggered by assuming Weak Realism, we can use the twisted Malus law, that gives us:
\begin{equation}\label{TMalus 2}
\langle \mathcal E, \mathcal P \rangle= -\cos (\theta_\mathcal E-\theta_\mathcal P)\,,
\end{equation}
to also obtain readily:
\begin{equation}\label{TMalus 3_1}
\langle \mathcal E, \mathcal P' \rangle= -\cos (\theta_\mathcal E-\theta_{\mathcal P'})
\end{equation}
and 
\begin{equation}\label{TMalus 3_2}
\langle \mathcal E', \mathcal P \rangle= -\cos (\theta_{\mathcal E'}-\theta_\mathcal P)\,.
\end{equation}
Let $\tilde{\mathcal Q}$ stand for the sequence or normalized spin projections along the angle $\theta_\mathcal Q$ but on the side opposite to the side corresponding to $\mathcal Q$.  Since in this subsection we are assuming Locality, we have the identity:
\begin{equation}\label{Conserv}
\tilde{\mathcal Q}_i+{\mathcal Q}_i\equiv 0
\end{equation}
for any  ${\mathcal Q}\in \{ \mathcal E, \mathcal P,  \mathcal E', \mathcal P'  \} $. The relation (\ref{Conserv}) is a direct consequence of the singlet state expression and Wave Packet Reduction if one at least of ${\mathcal Q}$ and $\tilde{\mathcal Q}$ is actually measured. For the other cases, one uses Locality to state that $\tilde{\mathcal Q}_i$ is unchanged if the setting is changed on the other side, where one could actually measure ${\mathcal Q}$.  But then, we notice that by Convention 1 and Locality, ${\mathcal Q}_i$ remains unchanged if it is measured instead of being inferred to make sense by invoking Weak Realism, so that in all cases the conclusion is the same as if one at least of ${\mathcal Q}$ and $\tilde{\mathcal Q}$ is observed. 

\medskip
We notice that if one does not assume Locality, then the identity (\ref{Conserv}) holds true when at least one of  ${\mathcal Q}$ and $\tilde{\mathcal Q}$ is actually observed, but not necessarily otherwise since one cannot then use the reasoning on which we relied to justify the relation (\ref{Conserv}) in the case when Locality is assumed to hold true. 

From equations(\ref{TMalus 3_1}) or (\ref{TMalus 3_2}) that are equivalent to each other by exchanging the sides of Alice and Bob,  we get readily:
\begin{equation}\label{TMalus 3_3}
\langle \mathcal E, \tilde{\mathcal E'}\rangle= -\cos (\theta_\mathcal E-\theta_{\mathcal E'})
\end{equation}
and 
\begin{equation}\label{TMalus 3_4}
\langle \tilde{\mathcal P'}, \mathcal P \rangle= -\cos (\theta_{\mathcal P'}-\theta_\mathcal P)\,,
\end{equation}
from which by (\ref{Conserv}) we respectively get:
\begin{equation}\label{TMalus 3_3_a}
\langle \mathcal E, {\mathcal E'}\rangle= \cos (\theta_\mathcal E-\theta_{\mathcal E'})
\end{equation}
and 
\begin{equation}\label{TMalus 3_4_a}
\langle {\mathcal P'}, \mathcal P \rangle= \cos (\theta_{\mathcal P'}-\theta_\mathcal P)\,.
\end{equation}
Using again (\ref{TMalus 3_3_a}), (\ref{TMalus 3_4_a}), and Locality, we also know that:
\begin{equation}\label{TMalus 3_5_a}
\langle \mathcal E',\tilde {\mathcal P'}\rangle= \cos (\theta_{\mathcal E'}-\theta_{\mathcal P'})
\end{equation}
and 
\begin{equation}\label{TMalus 3_5_b}
\langle {\mathcal P'},\tilde{\mathcal E'} \rangle= \cos (\theta_{\mathcal P'}-\theta_{\mathcal E'} )\,.
\end{equation}
Any of these two equations lets us compute $\langle {\mathcal P'},{\mathcal E'} \rangle$ as:
\begin{equation}\label{TMalus 3_6}
\langle \mathcal E',{\mathcal P'}\rangle= -\cos (\theta_{\mathcal E'}-\theta_{\mathcal P'})\,.
\end{equation}

We now have the values, hence also in particular the convergence of the finite sums as the numbers $N_\kappa$ diverge, for all the correlations that we need in both versions $V3$ and $V4$. Thus both of the Bell's Inequalities, \textit{i.e.,} equations (\ref{Sica_3p_4}) and (\ref{Sica_4p_4}), that we have formally deduced assuming convergence are fully justified and convergence if proved (and thus does not need to be assumed) if one assumes Weak Realism and Locality. In order to get from Bell's Inequalities to Bell's Theorem one needs to falsify at least one of these inequalities by choosing appropriate values of the parameters (the oriented axes or equivalently the angles). We will provide falsifications for both versions $V3$ and $V4$.

\medskip
- For version $V3$ we choose $\theta_{\mathcal P}=0$, $\theta_{\mathcal E}=\frac{3\pi}{4}$, and   $\theta_{\mathcal E'}=\frac{-3\pi}{4}$ so that  $\theta_{\mathcal E}$ and  $\theta_{\mathcal E'}$ differ by a right angle.  Since using Locality we easily get $\langle \mathcal E,\mathcal E'\rangle=0$, by further using $\langle \mathcal E,\mathcal P\rangle =\langle \mathcal E',\mathcal P\rangle=\frac{\sqrt{2}}{2}$, and by replacing all the correlations in equation (\ref{Sica_3p_4}) by their respective values we end up deducing the false inequality $\sqrt{2}<1$ by specialization of the $V3$ version of Bell's Inequalities. We can thus conclude that at least one of the assumptions that we have made,  Weak Realism and Locality, must be a violation of the laws of microphysics. Many other choices of angles would also work to generate a falsification of equation (\ref{Sica_3p_4}) or another Bell's Inequality. \textbf{Q.E.D.}

\medskip
- For version $V4$ we choose $\theta_{\mathcal E}=\frac{\pi}{4}$, $\theta_{\mathcal E'}=\frac{3\pi}{4}$, $\theta_{\mathcal P}=\frac{\pi}{2}$, and $\theta_{\mathcal P'}=0$, thus angular differences $| \theta_{ \mathcal E}- \theta_{ \mathcal P}|=| \theta_{ \mathcal E}-\theta_{ \mathcal P'}|=| \theta_{ \mathcal E'}-\theta_{ \mathcal P}|=\frac{\pi}{4}$ and $|\theta_{  \mathcal E'}-\theta_{ \mathcal P'}| =\frac{3\pi}{4}$.  Replacing the correlations in equation  (\ref{Sica_4p_4}) by their respective values we end up having deduced the false inequality $2\sqrt{2}\leq 2$ by specialization of the $V4$ version of Bell's Inequalities. Thus we can  again conclude that at least one of the assumptions that we have made,  Weak Realism and Locality, must be a violations of the laws of microphysics. Many other choices of angles would also work here, but the example chosen here for $V4$ is optimal in terms of the worse falsification of (\ref{Sica_4p_4}).  \textbf{Q.E.D.}

\medskip
\noindent
\textbf{Remark 1.} \emph{As Sica noticed in \cite{Sica1}, the finite $N_\kappa$ equations  (\ref{Sica_3p_3}) and (\ref{Sica_4p_3}) are identities, independently of any convergence property.  Sica calls them \emph{``Bell Identities"} to distinguish them form the \emph{``Bell Inequalities"} that follow from these identities if convergence hold true for all the averages.  The Bell Identities have to be satisfied as soon as one assumes Weak Realism that provides us the three or four sequences of $-1$'s and $1$ that are needed depending on which of these two identities we want to work with. Nevertheless, it would at best hard to \textbf{only} use the Bell Identities to get a contradiction if one had no proof of convergence whatsoever since then one would not have means to evaluate the terms in the identities (whether one deals with finite sums or with their asymptotic versions).} 

\bigskip
\noindent
\textbf{3)}  \textbf{A Bell's Theorem with no Locality assumption.}
In subsection 2.2, we have recalled the classical theory of Bell, in two versions $V3$ and $V4$ where the number in the name of the version is the number of oriented axes used to obtain normalized projections of the spins. We completed this task assuming  Weak Realism and Locality. 

\medskip
\emph{But what happens if Locality is replaced by the EACP?} 

\medskip
\noindent
After examining the new hypothesis and its strength when compared to Locality, we will first investigate what remains of the computability of the various correlations that are related by one or another Bell Inequality. We will easily deduce from this investigation that only $V3$ can possibly be dealt with when assuming the EACP instead of Locality.  This is because one of the correlations in equation (\ref{Sica_4p_4}) cannot be evaluated, nor even guaranteed convergence with the substitute hypothesis. But before that, we need to make sure that we are not using the same hypothesis under disguise.

\bigskip
\noindent
\textbf{3.1:}  \textbf{Statement of the EACP vs Locality Lemma.}
We will use, here and in the next subsection, a specification of the Effect After Cause Principle that is quite focused on the entities that we deal with in Bell's Theories, but the reader is advised that warnings [W1]-[W3] are still in vigor and are essential for a correct understanding of any of the definition of the EACP

\bigskip
\noindent
\textbf {Effect After Cause Principle (\emph{EACP: specific form}).} \emph{For any Lorentz observer and for any $\mathcal Q$ in $\{ \mathcal E,\, \mathcal E',\, \mathcal P ,\, \mathcal P'  \}$, a value $\mathcal Q_i$ of $\mathcal Q$ cannot change as a result of a cause that happens after $\mathcal Q_i$ has been measured for that observer, or should already exist with a definite value in the case when $\mathcal Q_i$ is not observed but results from the chosen form of Realism,.}

\medskip
This version of the EACP adapted to the context of Bell's Theory will be used to prove the following comparison result that is crucial to our purpose.

\medskip
\noindent
\textbf{EACP vs Locality Lemma.} \emph{The EACP is different from Locality and not stronger than Locality, meaning that the EACP does not imply Locality.}

\emph{Proof of  the EACP vs Locality Lemma.} We assume Weak Realism and the EACP and notice that, in view of Warning [W1]-[W3], the EACP has been formulated so as to be compatible either with Locality or with Non-Locality, whichever one chooses.  \textbf{Q.E.D.} 

\smallskip
\noindent
\textbf{EACP weaker than Locality.} \emph{ Furthermore,  the EACP is weaker than Locality. Otherwise speaking, Locality implies the EACP but the reverse implication is not true.}

\smallskip
\noindent
\emph{About the EACP weaker than Locality statement.}  The relative strength statement follows from the easily checked fact that the failure of the EACP permits \emph{Superluminal Message Transmission} (or SMT), using readable messages (the ones we all know how to use) or at least (depending on how the EACP fails to hold true) \emph{Realist Superluminal Message Transmission} (or RSMT), \textit{i.e.,} a form of SMT using messages that are collections of values some or all of which only exist because of the chosen Realism assumption.  Since one knows that the violation of Locality does not permit SMT (see, \textit{e.g.,} \cite{Eberhard1978, GRW1980, Jordan1983}), we know that Locality and \emph{Causality} (the impossibility of SMT) do not coincide in a world without augmentation of Quantum Mechanics by any form of realism and that indeed Locality is stronger that Causality  in such a world. Furthermore Locality is stronger that Causality extended to be the impossibility of both SMT and RSMT in any world since we assume the impossibility of RSMT. We deduce that the EACP is weaker than Locality in a world without augmentation of Quantum Mechanics by any form of Realism, and more generally as long  (at least) as one considers RSMT to be impossible.  In fact, we consider RSMT to be as impossible as standard SMT, independently of Weak Realism being perhaps not compatible with the actual laws of Physics, considering that \emph{an extension of Physics by variables that are not only inaccessible but also violate fundamental laws obeyed by regular variable basically escape the Realm of Physics}.  Hidden Variables, one of the strong forms of Microscopic Realism, are unaccessible but otherwise obey the law of Physics, and many who are ready to support Non Locality rely for that on the fact that Non Locality does not permit SMT: it does not seem compatible with Physics to allow RSMT on the sole basis that it is all about things out of our reach. Details are left to the reader.  

The use of the symbol  \textbf{Q.E.D.} for the above discussion is perhaps justified and some could consider the \emph{EACP weaker than Locality statement} as a Lemma just as good as the \emph{EACP vs Locality Lemma}. We chose to avoid a polemic on this point since the \emph{EACP vs Locality Lemma} is strong enough to make our Bell Theorem different from the usual one.  Furthermore many would consider the EACP as a minimal assumption to stay in the Realm of Physics.  This does not prevent that because of what is not surely established about RSMT, the ``proof" that we provide is even formally a physicist argument and not any more a mathematical proof.  Such an argument in the form of a discussion rather than a proof is not the last one we will provide since we will meet RSMT again below in the discussion of the \emph{No-Correlation Lemma}.  But should one be surprised that Physics does not reduce to a mathematical discussion anyway?

\medskip
\noindent
\textbf{Remark 2.} \emph{Comparing now the EACP and Locality in terms of what are the correlations that make sense and can be computed assuming only Weak Realism as a supplementary hypothesis, we see that the lists of computable correlations are quite different. In particular, because Non-Locality is not assumed conjointly to the EACP, the correlation  $\langle \mathcal E',\mathcal P'\rangle$ does not make sense as long as one does not \textbf{also} assume Locality or anyway some strong enough extra condition on top of the EACP.  It follows that version $V4$ of the Bell inequality cannot be used if one replaces assuming Locality by assuming the EACP only. The status of $V3$ is different enough and we will come back to the problem of the computation of  $\langle \mathcal E,\mathcal E'\rangle$ or $\langle \mathcal P,\mathcal P'\rangle$ under the EACP assumption in some particular cases in the proof of what we call The No-Correlation Lemma below.  To the contrary, we notice that formulas \ref{TMalus 2} to \ref{TMalus 3_2}, together with enough of the convergence properties embedded in these formulas, remain valid when assuming only the EACP as Augmentation and Weak Realism to give meaning to the various factors.  These differences about which correlation functions that can be computed, or simply  make sense, may be considered as an alternate proof of the EACP vs Locality Lemma, but we notice that the main ingredient of this comparison is again the compatibility of the EACP with both Locality and Non-Locality, whichever one chooses as extra hypothesis about the independence or dependence of measurement upon the setting of a spatially remote apparatus setting.  See also Remark 2 in subsection 3.2.}
  
\bigskip
\noindent
\textbf{3.2:}  \textbf{The No-Correlation Lemma.}

\smallskip
\noindent
{\bf No Correlation Lemma.} \emph{Assuming the EACP, if the oriented axes $a_{\mathcal E}$ and $a_{\mathcal E'}$ are orthogonal to each other, then the sequences $\mathcal E$  and $\mathcal E'$ are not correlated,} and more precisely we have:
\[
(\circ)\quad  \langle\mathcal E ,   \mathcal E'  \rangle=0 \quad\textrm{or equivalently}\quad Prob(\mathcal E_i = \mathcal E'_i)=\frac{1}{2}\,.
\] 
or more generally 
\begin{equation}\label{liminf}
\liminf_{N\to\infty}\frac{1}{N}(\mathcal E_{i+1}\cdot \mathcal E'_{i+1}+\dots
+\mathcal E_{i+N}\cdot \mathcal E'_{i+N})\leq 0\,,
\end{equation}
and 
\begin{equation}\label{limsup}
\limsup_{N\to\infty}\frac{1}{N}(\mathcal E_{i+1}\cdot \mathcal E'_{i+1}+\dots
+\mathcal E_{i+N}\cdot \mathcal E'_{i+N})\geq 0\,.
\end{equation}

\noindent

\medskip
Before going further, we give a definition that will be useful whenever dealing with the EACP throughout the rest of the paper.

\medskip
\noindent
\textbf{Definition 2.}  \emph{With $(X,Y)\in \{(E,P), (P,E)\}$ an \emph{$X$-$Y$ observer} is a Lorentz observer for whom measurements at the measurement tool $X$ occur before measurements at measurement tool $Y$ for each pair produced at $S$.}  

\emph{Proof of the No Correlation Lemma.}
We will first prove an auxiliary result (the \emph{Restricted No Correlation Lemma}) that corresponds to about the same statement, but when no measurement is made on the $P$ side. 
Then we will show that for an $E$-$P$ observer, making a measurement or not at $P$ may violate the EACP so that if a measurement made at $P$ can change the correlation on the $E$ side, one has a violation of the Free Will Principle. 

\smallskip
\noindent
{\bf Restricted No Correlation Lemma.} \emph{Assuming the EACP and assuming further that no measurement is made on the $P$ side, if the oriented axes $a_{\mathcal E}$ and $a_{\mathcal E'}$ are orthogonal to each other, then the sequences $\mathcal E$  and $\mathcal E'$ are not correlated, and more precisely we have:
\[
(R-\circ)\quad  \langle\mathcal E ,   \mathcal E'  \rangle=0 \quad\textrm{or equivalently}\quad Prob(\mathcal E_i = \mathcal E'_i)=\frac{1}{2}\,.
\] 
or more generally (\ref{liminf}) and (\ref{limsup}). }

\medskip
\noindent
\emph{Proof of the Restricted No Correlation Lemma.}
Using the EACP and the conclusions that can readily be deduced from it, and more precisely the fact that ``the three sequences $\mathcal E$, $\mathcal E'$, and $\mathcal P$ involved in equation (\ref{Sica_3p_4})  are well defined",  we notice that, if furthermore no measurement is made on the $P$ side, then  only the orientation of the angle  $< a_{\mathcal E};  a_{\mathcal E'}>$ at $E$ could matter for an $E$-$P$ observer, so that $(R-\circ)$ follows from invariance under Parity without assuming Locality if one assumes convergence of the means that define the correlations that are of intetrest to us.  Otherwise, when one cannot prove convergence, the argument that we have given proves the inequalities (\ref{liminf}) and (\ref{limsup}).  We next provide some details that some readers may prefer to avoid.

To see the role of Parity, we introduce the further oriented axis $ a_{\mathcal E''}$ that is parallel to $ a_{\mathcal E'}$ but with the opposite orientation.  This is the (only) oriented axis to which would correspond the sequence $\mathcal E''$ such that  $\mathcal E''_i\equiv -\mathcal E'_i$.  Since 
\begin{equation}\label{Sum2One}
Prob(\mathcal E_i = \mathcal E'_i)+Prob(\mathcal E_i = \mathcal E''_i)=1\,,
\end{equation}
it only remains to prove that these two probabilities are equal to each other.  We use here sequences whose values are possibly unknown (and indeed forever inaccessible to our knowledge), but that are  known to be well defined as we have recalled to begin this proof:

- One of these sequences, $\mathcal E$, is known by direct measurement, 

- The other sequence, $\mathcal E'$,  can be inferred to be well defined, even if unknown, by an $E$-$P$ observer on the basis of Quantum Mechanics augmented by  Weak Realism. 

\noindent
Since the angles $< a_{\mathcal E};  a_{\mathcal E'}>$ and $< a_{\mathcal E''};  a_{\mathcal E}>$ are equal, using the EACP, the only thing that could generate an inequality between $Prob(\mathcal E_i = \mathcal E'_i)$ and $Prob(\mathcal E_i = \mathcal E''_i)$ for an $E$-$P$ observer is the difference in the orientations of the angles $< a_{\mathcal E};  a_{\mathcal E'}>$ and $< a_{\mathcal E};  a_{\mathcal E''}>$.  Equality thus follows from Parity invariance if one assumes convergence of the means that define the correlations that are of intetrest to us.  Otherwise, in full generality, the argument that we have given proves the inequalities (\ref{liminf}) and (\ref{limsup}). 
  \textbf{Q.E.D.}

Coming back now to the full No Correlation Lemma, we see that the proof given for the Restricted No Correlation Lemma cannot work if one does not assume Locality as the possible dependence of $E$ and $E'$ upon the choice of $P$ breaks the symmetry between $< a_{\mathcal E};  a_{\mathcal E'}>$ and  $< a_{\mathcal E'};  a_{\mathcal E''}>$.  This is why we need to deal with the fact that measurements along one axis will indeed be made on the $P$ side: more precisely such measurements may be made to study the full version of the  No Correlation Lemma, and then (for other samples) need to be done to write the desired Bell's Inequality.   

Thus assume that some long sequence of pairs are emitted and measured in a very short time, enough to have obtained what could look as an asymptotic value, say $v\leq0$ for the correlation $\langle \mathcal E_{i}, \mathcal E'_{i}\rangle$ (a value unknown but that exists as a consequence of Weak Realism). Then if the measurement or not at $P$ can change this value $v\leq 0$, and in particular change it to another value $v'>0$, then one would have a contradiction of the EACP for an $E$-$P$ observer, so that we can conclude that the Free Will Principle imposes that making measurements at $P$ cannot change the value of the correlation values (or more generally the asymptotic bounds on correlation) that we arrived at in the Restricted No Correlation Lemma. This finishes the proof of the No-Correlation Lemma.  \textbf{Q.E.D.}

We recall that we have seen that values at $E$ or $E'$ by themselves do not create a conflict with the Free Will Principle, that becomes sensitive only in the case that we have isolated and treated in the Restricted No Correlation Lemma. 

We also notice that an entity able to access the values that exist in virtue of Weak Realism besides the values that can be measured would be capable of RSMT if the correlation would be changed when measurements are made on the $P$ side. This remark, using again RSMT as in the comparison of the EACP with Locality, could possibly lead to an alternate proof of the No Correlation Lemma that would not need the Free Will Principle. As in the comparison of EACP with Locality, a formal version of such a proof eludes us because we can only prove  that RSMT should be considered just as impossible as usual SMT using a strong appeal to an opinion. Many physicists including this author would not mind making some call on opinion and would rather consider that banishing opinions from Physics is probably a misunderstanding of what science is all about. Everyone would nevertheless agree that opinions should be minimized and as much as possible justified which is why we are happy to have more formal proofs of the most delicate points to offer here in complement to what was offered in \cite{TresserAutre}.

At last (as an ultimate defense of of he genuine Bell's Theorem part of  \cite{TresserAutre}), we want to point out that the breaking of the symmetry  between $< a_{\mathcal E};  a_{\mathcal E'}>$ and  $< a_{\mathcal E'};  a_{\mathcal E''}>$ by possible measurements at $P$ requires a particularly strong version of Non-Locality, and more precisely one that can break remotely the Parity symmetry.  As a consequence, a former version of this result \cite{TresserAutre} that, without these words, used the proof of the Restricted No Correlation Lemma as a proof of the full version of the No Correlation Lemma may probably be considered as valid by a part at least of the community of physicists.  The care given here of the difference between the Restricted and the full version will hopefully convince a significant proportions of those who at least would have had some doubts about that former version.

\bigskip
\noindent
\textbf{3.3:}  \textbf{A Bell's Theorem with no Locality assumption.}
The tools and fact that we have assembled let us formulate and prove the following Main Result:

\smallskip
\noindent
\textbf{New Bell's Theorem:} \emph{Assuming  Weak Realism and the EACP, we can use the triplet
of angles $(\theta_\mathcal P,\theta_\mathcal E,\theta_{\mathcal E'})= (0,\frac{3\pi}{4}, \frac{-3\pi}{4})$
that corresponds to the triplet of correlations
$( \langle  \mathcal P , \mathcal E\rangle,\langle\mathcal E' ,\mathcal P\rangle,  \langle\mathcal E , \mathcal E'\rangle)=(\frac{\sqrt{2}}{2}, \frac{\sqrt{2}}{2}, 0)\,$  
to generate a contradiction using the $V3$ version of Bell's Inequalities.}

\smallskip
\emph{Proof of the New Bell's Theorem.}  Again we assume the EACP and we use
equation (\ref{Sica_3p_4}) as the inequality to be falsified.
 
 \smallskip
- 1) After measurements are made using $P$, a $P$-$E$ observer obtains  that:

\noindent
- e1) $\langle \mathcal P,\mathcal E\rangle=\frac{\sqrt{2}}{2}$, \textit{i.e.,}  $\langle \mathcal P,\mathcal E\rangle\approx 0.7$ by Quantum Mechanics, or by direct observation after measurements are also made using $E$,

\noindent
 - e2) $\langle \mathcal P,\mathcal E'\rangle= \frac{\sqrt{2}}{2}$, \textit{i.e.,} $\langle \mathcal P,\mathcal E'\rangle\approx 0.7$ by Quantum Mechanics augmented by  Weak Realism.
 
 The deductions made in e1) and e2) using Quantum Mechanics augmented by Weak Realism go as follows:  By Wave Packet Reduction (for instance), the spin state of second particle (the particle on the $E$ side) becomes 
\begin{equation}\label{Collapse}
\Psi(x_2)=|\mathcal P_i \rangle _1\otimes| -\mathcal P_i \rangle_2 
\end{equation}
along the oriented axis along which the sequence $\mathcal P$ is measured, as soon as the measurement of $\mathcal P_i$ is made on the $P$ side.  Hence the second particle gets into a spin state prepared to be $ |-\mathcal P_i \rangle $ along that oriented axis
(as revealed by using the information obtained on the $P$ side) so that both of the two correlations $\langle\mathcal P,\mathcal E\rangle$ and $\langle \mathcal P,\mathcal E'\rangle$ are equal to $\frac{\sqrt{2}}{2}$ (about $0.7$) by a simple application of the twisted Malus law as we have recalled it, under Convention 2 and the EACP, as we saw in Remark 2.

 \smallskip
- 2) An $E$-$P$ observer infers that: 

\noindent
 - e3) $Prob (\mathcal E_i=\mathcal E'_i)= 0.5$ (\textit{i.e.,} $\langle \mathcal E, \mathcal E'\rangle =0$) on the $E$ side by the No Correlation Lemma or at least the pair of inequalities \ref{liminf} and \ref{limsup} whose utilization (in fact we would only use \ref{liminf} on course)  would force us to use some specific times in order to falsify the chosen Bell Inequality, and more precisely times such that the value of the infimum limit in \ref{liminf} is well enough approximated and in particular such the partial sum is negative. Such times should also be long enough and in particular chosen so that the other two correlations have values that are good enough approximations of the asymptotic value:  no problem of incompatibility in the choice of the time can occur since  the averages that define $\langle \mathcal P, \mathcal E\rangle$ and $\langle \mathcal P, \mathcal E'\rangle$ both converge as we saw in e1) and e2). 

\smallskip
Assembling the conclusions  e1), e2), and e3) from the two (strongly) asynchronous frames (\textit{e.g.,} in the Lorentz frame of the experiment since the outcomes cannot change according to the Lorentz frame by relativistic invariance of observable events) one obtains  the expected triplet evaluation for the three correlations: $(\frac{\sqrt{2}}{2}, \frac{\sqrt{2}}{2}, 0)$.  Together with equation (\ref{Sica_3p_4})  (the $V3$ version of Bell's Inequalities), this evaluation provides us with the impossible inequality $1.4\leq 1$, or more precisely $\sqrt{2}\leq 1$ as when we examined equation (\ref{Sica_3p_4}) while assuming Locality in Subsection 2.2.  This concludes the proof of the New Bell's Theorem.  \textbf{Q.E.D.}

\medskip
Our New Bell's Theorem admits the following immediate corollary that we will use as our main conclusion:

\medskip
\noindent
\textbf{ Conclusive Corollary:} \emph{ Weak Realism is the \textbf{only} possible cause of contradiction common to \textbf{all} the versions of Bell's Theorem and some of the Bell's type contradictions cannot be solved by assuming Non-Locality. Thus Non-Locality is not needed (in some circles, one would say that Non-Locality can be disposed of using Occam's razor).} 

\medskip
\noindent
\textbf{Remark 3.} \emph{As we saw, without Weak Realism any violation of the EACP is a violation of Causality since then the EACP is one of the expressions of Causality.  In order for the violation of the EACP to not be a violation of Causality, one would have to accept that the negation of the EACP has effect \emph{only} on values of observable that are linked to Weak Realism.  We do find that unacceptable and we thus consider that the new Bell Theorem condemns Weak Realism. This is not a proof and possibly no actual proof can be given to help us decide between keeping Weak Realism and keeping the EACP but this situation is more frequent than one might think.  Mathematics and not Physics is the Realm of ``proofs": there has always been some opinions lurking behind the way we apprehend the laws of Physics. Some would say that as soon as one uses  Weak Realism, one steps into Metaphysics anyway.  However we would prefer an argument a bit more subtle than just stamping the word ``philosophy" on anything that uses any form of Realism. This is why we point out that  Weak Realism violates the spirit if not the letter of the Uncertainty Principle \cite{Heisenberg} or at least its time-reversed version \cite{ETP, Tresser2}, and is in particular rejected by the Copenhagen Interpretation of Quantum Mechanics (which consequently would have to statute that Bell's Theorem has nothing to say about Quantum Mechanics).  On the other hand, invoking  Weak Realism, even if bad, is probably not as bad as accepting that the EACP is false.  Yet the present work shows that accepting the mildly unacceptable Weak Realism implies accepting the quite unacceptable violation of the EACP.}

\medskip
\noindent
\textbf{Remark 4.} \emph{The well known Hidden Variable theory of de Broglie and Bohm \cite {deBroglie19271956}, \cite{Bohm1952} is both Non-Local and Realist, yet it avoids the contradictions that constitutes BellÕs theorem.  In fact it avoids these contradictions  precisely because Non-Locality prevents the Bell's inequality from making sense. This statement on the de Broglie - Bohm Theory (dBBT) is not a contradiction to our Conclusive Corollary about Weak Realism and Locality nor more generally to the theses defended here.  Indeed, the dBBT is massively not Lorentz invariant, way beyond the special setting for Bell's theory, and apparently irreducibly so (see in particular \cite{FreeWill1} and references therein): in any Bohmian quantum theory the quantum equilibrium distribution $|\Psi|^2$ cannot simultaneously be realized in all Lorentz frames of reference.  To the contrary, Quantum Mechanics can be viewed as a non-relativistic approximation to relativistic Quantum Field Theory in the limit when Classical Mechanics is a good approximation to Special Relativity. The dBBT, or Bohmian mechanics, the version re-discovered and extended by Bohm, is thus false or at least considered as false by most physicists, even if it can serve pedagogically as advocated by Bell (this opinion of Bell, who defended Bohmian Mechanics, may not be shared by those who consider Non-Realism as an essential ingredient of microphysics).  Indeed, some  Bohmian physicists still hope for a version of Bohmina Mechanics that could be acceptable by the profession, but it should be noted that (many) Bohmian physicists take as a strong argument in favor of Bohmian mechanics the false statement that \emph{Bell's Theorem and Alain Aspect's experiments and similar ones prove Quantum Mechanics to be a non-local theory}. Indeed, in some sense, Bell's Theorem can be considered as the proof that \emph{the Non-Local character of the dBBT was irreducible among Hidden Variable Theories and more generally among Realist Theories}.  To the contrary since it assumes (Weak or stronger) Realism, hence forces us out of Quantum Mechanics (into Bohmian Mechanics or some weak form of it), \emph{Bell's Theory and related experiments have nothing to say about the Locality or non-Locality of Quantum Mechanics itself}. This being said, Bell and others consider that the complete nature of Quantum Mechanics can be trusted as a consequence of what the EPR paper proves Quantum Mechanics itself to be non local, but here is not the place for a debate on the History of Sciences.}

%%%%%%%%%%%%%%%
%%%%%%%%%%%%%%
%%%%%%%%%%%%
%%%%%%%%%
%%%%%
%%%
%%
%
\medskip
Like in the case of the EPRB entanglement (see \textit{e.g.,} \cite{AspectEtAl1982}), experiments have been done on other entanglements (see \textit{e.g.,}  \cite{BPWZ1999}).  Such other entanglements, that comprise the GHZ entanglement (see \cite{GHZ1989, Mermin1990GHZ3, GHSZ1990GHZ3} and also textbooks such as, \textit{e.g.,} \cite{Peres1993} or \cite{Le Bellac} that also cover EPR entanglements) and Hardy's entanglement (see \cite{Hardy1993}) will not be considered here, but we recall that the GHZ entanglement (with the associated \emph{``Bell Theorem without inequalities"}) is treated in \cite{TresserAutre} where Locality is replaced by the EACP as for the genuine Bell's Theorem.

\bigskip

\begin{acknowledgments}
\textbf{Acknowledgments:} 
Many people have helped me in this enterprise (the conjunction of the present paper and \cite{TresserAutre}), by vicious attacks, constructive questions, encouragements,  patient listening, and often friendship: Y. Avron, M. le Bellac, O. Cohen, P. Coullet, D. Greenberger, R. Griffiths, G. t'Hooft, A. Mann, D. Mermin, D. Ostrowsky,  I. Pitowsky, Y. Pomeau, O. Regev, M. Revzen, T. Sleator, J. Tredicce,  L. Vaidman and many more.  Some people could have as well been on the front page but declined.  I cannot find the words to thank M. E. Brachet, A. Fine, R. Friedberg, P. Hohenberg, L. Horwitz, M. Martens, and E. Spiegel for their patience, strong but important and legitimate critics, advices, and encouragements.  Any part of this paper could be claimed \emph{in particular} by Richard Friedberg whose help has been considerable, and often crucial. Nevertheless, any error is mine.
\end{acknowledgments}


\end{document}